%% file: shadow4.tex
\newif\iflayout
\layouttrue

\iflayout
   \documentstyle[cite,aps,epsfig,multicol,fancy,color]{revtex}
   \tightenlines	
\else
   \documentstyle[preprint,eqsecnum,aps]{revtex}
\fi

\iflayout
\else
   \draft
\fi

\begin{document}
\newcommand{\bea}{\begin{equation}}
\newcommand{\eea}{\end{equation}}
\newcommand{\ber}{\begin{eqnarray}}
\newcommand{\eer}{\end{eqnarray}}

\draft \preprint{\today}
\twocolumn[\hsize\textwidth\columnwidth\hsize\csname
@twocolumnfalse\endcsname

\title{A Thermal Re-emission Model}
\author{${}^{1,2}$ Amit K. Chattopadhyay$ {}^{*} $}
\address{${}^{1}$ Department of Chemistry, University of Virginia, McCormick Road, Charlottesville, VA 22904, U.S.A. \\
${}^{2}$ COMP/Laboratory of Physics, Helsinki University of Technology, P.O.Box 1100, FIN-02015 HUT, Finland}

\maketitle
\iflayout
 \thispagestyle{fancy}
\fi

\begin{abstract}                
Starting from a continuum description, we study the non-equilibrium roughening of a thermal re-emission model for etching in one and two spatial dimensions.
Using standard analytical techniques, we map our problem to a generalized version of an earlier non-local KPZ
(Kardar-Parisi-Zhang) model. In 2+1 dimensions, the values of the roughness and the dynamic exponents calculated from our theory go like
$ \alpha \approx z \approx 1
$ and in 1+1 dimensions, the exponents resemble the KPZ values for
low vapor pressure, supporting experimental results. Interestingly, Galilean invariance is maintained althrough.
\end{abstract}

\pacs{68.35.Ct, 05.40. -a, 05.70.Ln, 64.60.Ht}

\vskip2pc]

\narrowtext 
The subject of kinetic roughening and 
non-equilibrium growths, have been in the
center of interest of far-from-equilibrium physics for more than
two decades now. This is mainly due to two reasons: on the one
hand, due to an ongoing revolution in the world of micro-physics
in recent years, the demand of the age is to understand and
implement the underlying mechanism associated [1]. On the other
hand, they seem to correlate fields even as diverse as 
ecological growths, propagation of a crack-front, stock-market
predictions, etc. [2]. Although the processes which have been probed so far, 
have mostly been concerned only with local effects, such as molecular-beam-epitaxy (MBE) growth, conventional
diffusive growths, etc, the importance of the non-local effects, have been known as early as the 1950's [3]. Later on, with the advent of more sophisticated
experimental techniques, non-linear effects involving  physical vapor deposition (PVC) [1,4,5,6], sputtering techniques and associated growth and
etching of plasma fonts have assumed a position of paramount
importance. Whereas in standard MBE type of growths, the vapor
atoms are targetted in a direction normal to the substrate, so
that growth is decided by the local environment only, in case of
shadowing growths by sputter deposition, vapor atoms are incident
at random angles to the surface, so that non-local factors gain
prominence in this case [7-11]. There have been several
experimental follow-ups too of this sputtering mechanism [12-14].

The concept of shadowing effect in a sputtering growth (or
etching) essentially arrived with the observation that thin films
often exhibit "an extended network of grooves and voids in their
interiors" [11] giving rise to columnar structures. The basic idea
is the following. Since, in a sputtering growth (etching),
particles are allowed to be deposited (deroded) on the surface
from all possible angles at random, the rate of growth is taken to
be proportional to the exposure angle $ \theta(x) $, which is a
function of the position of incidence of the incoming particle.
Now, as the hills have greater exposure area, they receive more
atoms than the valleys. Thus the hills continue to grow steeper
compared to the depleted valleys, which naturally gives rise to an
instability in the system. The idea has been very ingeniously, but
intelligently related to the growth of the relatively larger
stalks, in a grassy lawn, which suppress the growth of the shorter
ones [11] and in the process giving rise to a rough contour. 

In the theoretical front, this phenomenon of shadowing growth
(decay), or its partner, the thermal reemission instability has inspired a series of works in 1+1
dimensions [7,8,11,15-17] and in 2+1 dimensions [14,18,19]. The
theoretical forays in fact started with the paper by Karunasiri,
et al [7] where from a direct numerical integration of the dynamical equation, they were able to show that the self-similarity of the contour, evident at small values of the diffusion constant, is modified by the growth of
flat films, beyond a critical height, as the value of the
diffusion constant is increased. Taking clues from their
arguments, Roland and Guo [15] went on to calculate the value of
the roughness constant, in 1+1 dimensions (albeit in the context of a shadowing model) and further predicted
that in the low temperature phase, the system resembles a KPZ
universality class (in agreement with Karunasiri, et al [7]). This
concept of non-local, shadowing effect 
was later modified in [9,11], where a net
non-local flux was observed to give rise to the inherent columnar
structures found in experiments. Later on, the domain of 2+1
dimension was also probed with the advent of advanced numerical
integration algorithms and Monte-Carlo simulations [18,19].
However, all these attempts, both in 1+1 and 2+1 dimensions, being
predominantly numerical, either through direct numerical
integration of a fundamental Langevin-type equation, or through
Monte-Carlo simulation, and all the more, giving contradictory
values of the exponents obtained by different groups, we ventured an analytical derivation to have a final say regarding the universality class of
these type of sputtered mechanisms. In the process, we will see
that our findings correlate the available experimental and
numerical observations (of one of these groups)  
in 2+1 dimensions and predicts scaling in 1+1 dimensions too. 

With the assumption that the shadowing effect provides the dominant instability in the system, we apply the non-local model proposed by Zhao, et al [14,18,19]. The model is given by

\bea
\frac{\partial h(\vec r,t)}{\partial t} = \nu {\nabla}^2
h(\vec r,t) \pm \sqrt{1+{({\vec \nabla}h)}^2}\:R(\vec r,t) +
\eta(\vec r,t)
\eea

{\noindent} 
and

\bea \langle \eta(\vec r,t) \eta(\vec r',t') \rangle =
2D\:{\delta}^2(\vec r-\vec r')\:\delta(t-t') \eea

{\noindent} where the first term on the right hand side of eqn.(1)
provides the diffusive relaxing mechanism for the growing (or
etching) surface and the last term signifies the collective effect
of randomness in the system, taken to be a Gaussian noise. The
middle term is the non-local, non-linear term detailing the
effects of thermal reemission and is given by

\bea R(\vec r,t) = s_0\:F_0(\vec r,t) + s_1\:F_1(\vec r,t) \eea

{\noindent} Here
$ s_0 $ is the
zeroth order sticking coefficient and 
$ s_1 $ is generated due to the reemission mechanism [14]. Here we consider first-order thermal reemission, that is neglect the effects of $ s_i $ $ \mathrm{(i>1)} $. 
Plugging again from the same reference and applying the same
logic, we consider the flux of the m-th order particle at position
$ \vec r $ as $ F_m(\vec r,t) $ which is given by

\ber
F_{m+1}(\vec r,t) &=& (1-s_m)\:\int Z(\vec r,\vec
r',t)\:F_m(\vec r',t) \times \nonumber \\
& & \frac{({\hat n_{{\vec r}{\vec r'}}}.{\hat
n})\:P({\hat n_{{\vec r'}{\vec r}}},{\hat n'})}{{(\vec r-\vec
r')}^2 + {(h-h')}^2}\:dA' 
\eer

{\noindent} 

For our case of first-order reemission, we are concerned with m=0 and 1. Here $ \hat n $ is the unit normal to the surface at $
\vec r $, pointing outwards, $ \hat n' $ is the unit normal at $
\vec r' $ and $ \hat n_{{\vec r}{\vec r'}} $ is the unit vector
connecting $ \vec r $ and $ \vec r' $ (see Fig.1). $ P(\hat
n_{{\vec r'}{\vec r}}, \hat n') $ is the probability, per unit
solid angle that the reemitted particle flies off along $ \hat
n_{{\vec r'}{\vec r}} $ and is expressed as $ \frac{(\hat n_{{\vec
r'}{\vec r}}.{\hat n'})}{\pi} $ [18]. $ Z(\vec r,\vec r',t) $
is
equal to unity except when there is no line of sight between the surface
elements at $ \vec r $ and $ \vec r' $ and zero otherwise. 
The nonlinear
factor $ \sqrt{1+{(\vec \nabla h)}^2} $ which is multiplied with $
R(\vec r,t) $, signifies the lateral growth (or etching, as the
case may be) associated and the '+' and '-' signs as its prefix,
refer to growth or etching respectively. In the following
analysis, we will consider parameter values as in [19] (that is we
will be dealing with etching due to sputtering). Thus, for our
case, $ F_0 = 4 $, $ s_0 \approx 0 $ and $ s_1 \approx 1 $. Also $
P(\hat n_{{\vec r'}{\vec r}},{\hat n'}) = \frac{\hat n_{{\vec
r'}{\vec r}}.{\hat n'}}{\pi} $, assuming thermally re-emitted
flux, although this is more of a simplification [20] than exact
truth. With the above description of the complete equation, we
proceed to determine the dynamics in the 2+1 dimensional case.
Later on, we will also discuss our results with reference to 1+1
dimensions, as well.

\begin{figure}
\centerline{\epsfig{figure=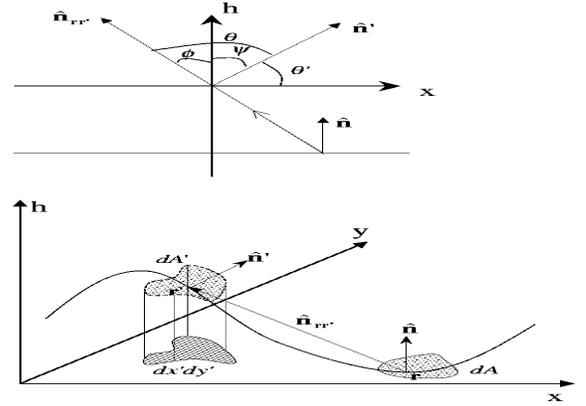,height=5.5cm,width=\linewidth}}
\caption{Relative orientations of the unit normals at $ \vec r $
and $ \vec r' $ and the co-ordinate system described by them.}
\label{Fig:1}
\end{figure}

\par 
Combining eqns.(1), (3) and (4) and
taking $ \psi $ as the angle between $ \vec r $ and $ \vec r' $
(see Fig.1), the dynamical etching equation reduces to

\bea 
\frac{\partial h}{\partial t} \approx \nu {\nabla}^2 h -
\lbrack{1 + \frac{1}{2} {(\vec \nabla h)}^2}\rbrack\:F_1(\vec r,t)
+ \eta(\vec r,t) 
\eea

{\noindent} 
where

\ber 
F_1(\vec r,t) &\approx& \int \int\:\frac{4 \cos{\theta}}{\pi}\:
\frac{\sin[\theta +{\theta}']}{{(\vec r-\vec r')}^2 +
{(h-h')}^2} \times \nonumber \\
& & \sqrt{1+{(\vec \nabla h'(\vec r',\theta'))}^2}\:r'
dr' d{\theta}' 
\eer

{\noindent} 
where $ \theta $ = angle between $ \hat n_{{\vec
r}{\vec r'}} $ and $ \hat n' $ = $ \phi + \psi $ as in Fig.1 and
$ {\theta}' $ is again defined as in Fig.1. 
In arriving at
eqns.(5) and (6), we have deliberately chosen $ \hat n $ as one of
the axes in the two dimensional plane, to simplify calculations. This can be done, since on the average this holds true.
Also the standard lateral growth assumption, $ |\vec \nabla h| < 
1 $ has been employed too. This $ F_1(\vec r,t) $ can be further
reduced to

\bea 
F_1(\vec r,t) \approx \frac{8{<\cos{\theta}>}^2}{\pi}\:\int_{-L}^{L}
dr'\:\frac{|r' - r| [1 + \frac{1}{2}{({\partial}_{r'}
h')}^2]}{{(r' - r)}^2 + {[h - h']}^2} 
\eea

{\noindent} 
where $ L $ is the size of the system. It is important to mention here 
that in deriving eqn.(7) from eqn.(6), we have used the mean-value theorem,
since $ \pi/2 - \delta < \theta' < \pi/2 + \delta $ ($ \delta $ is an angular
strip around h), the range being evident from Fig.1. The "$\approx$" sign justifies the
fact that we have taken a mean-valued average, represented by "$<>$"around the h-axis, thereby
removing $<\cos{\theta}>$ outside the integral as a first-order approximation.
 Simplifying
further, we arrive at the analytically tractable form of $ F_1(\vec
r,t) $, as given below:

\bea 
F_1(\vec r,t) \approx \frac{8{<\cos{\theta}>}^2}{\pi}\:\int_{-L}^{L}
dr' \frac{[1 - \frac{1}{2}{({\partial}_{r'} h')}^2]}{|r' - r|}
\eea

{\noindent} 
In arriving at the above equations, we have put on a
very standard assumption for any non-local model that the height
difference $ (h-h') $, calculated between any two points $ \vec r
$ and $ \vec r' $ of the growing surface should be much smaller
than their distance of separation, {\it i.e.} $ |h-h'| \ll |\vec
r-\vec r'| $, a basic property expected of any non-local process.

With this assumption and the mean-valued average done beforehand, the 
equation of motion now becomes

\ber 
\frac{\partial h}{\partial t} &\approx& \nu {\nabla}^2 h - 
\frac{8{<\cos{\theta}>}^2}{\pi}\:\int_{-L}^{L}
dr'\:\frac{1}{|r'-r|}\:[1 + \frac{1}{2}{(\vec \nabla h)}^2]
\nonumber \\ &+& \frac{4{<\cos{\theta}>}^2}{\pi}\:\int_{-L}^{L} dr'
\frac{1}{|r'-r|}\:{({\partial}_{r'} h(r'))}^2 + \eta(\vec r,t)
\eer

{\noindent} 
Now, we try to look at the possible large time, long
distance behavior of the system. We can easily see that the KPZ
part [21], constituting the second term on the R.H.S. of the above
equation will vanish as the system size is taken to be
sufficiently large. In deriving the above form, terms higher than
$ {(\vec \nabla h)}^2 $ order have been neglected. The final
equation now looks like

\bea 
\frac{\partial h}{\partial t} \approx \nu {\nabla}^2 h + 
\int_{0}^{L} dr' \phi(r,r')\:{|{\partial}_{r'} h|}^2 + \eta(\vec
r,t) 
\eea

{\noindent} 
where

\bea 
\phi(r,r') = \frac{4\lambda \pi}{|r'-r|} 
\eea

{\noindent} 
$ \lambda = {\lambda}_0 \: {<\cos{\theta}>}^2 $
is an adjustable coupling parameter, such that we will later put $
{\lambda}_0 $ equal to unity. The fact that the assumptions employed above are perfectly trustworthy, can be cross-checked from the fact that eqn.(11) maintains translational invariance which was an important feature of our starting eqn.(4).

Eqn.(10) can be easily mapped to the phenomenological equation considered in [22]. The only trick lies in a suitable wave-vector representation of
the effective long-range potential $ \phi(r,r') $ in our case.
Obviously, this cannot be a simple plug-in from the earlier
equation of motion [22], since, here, the interacting potential is
apparently a multi-valued function. To progress further, we move
on to the wave-vector representation
of this interacting potential which is given by the scaled relation

\bea 
\phi(k,k') = 4\frac{\lambda}{k}\:f(\frac{k}{k'}) 
\eea

{\noindent} 
Here the scaling function looks like

\bea 
f(\frac{k}{k'}) = \int dX\: X e^{-iX}\:\int \frac{Y
e^{-iY}}{(Y - \frac{k}{k'} X)} 
\eea

{\noindent} 
Considering the scaling ansatz

\bea 
f(k,k') = f(\frac{k}{k'}) = A\:{(\frac{k}{k'})}^{\eta}, 
\eea

{\noindent} 
we get

\bea 
\phi(k,k') \approx \lambda \frac{k^{\eta-1}}{{k'}^{\eta}} 
\eea

{\noindent} 
and our job now is to evaluate the definite scaling
behavior for $ f(k,k') $ by the evaluation of a number for $ \eta
$ from eqn.(13) [24]. Applying simple Laplace transform and going
through the standard steps, it is easy to see that the dominating
contribution of the double integral in eqn. (13) implies that $
\eta = 1 $ [23] and this gives the value

\bea 
\phi(k,k') \approx \lambda \frac{1}{k'} 
\eea

{\noindent} 
{\it i. e.} the major contributing part of the
potential is effectively reduced to a single variable mode. Now,
we can simply plug-in results from ref.[22] and write down the
dynamic exponent $ z $ as

\bea 
z = 2 + K 
\eea

{\noindent} 
where

\bea 
K = -24/23 = -1.04 
\eea

{\noindent} 
for our case [24]. One obvious point to be noted here
is the fact that owing to the Galilean invariance of eqn.(9), we
can easily see that

\bea 
\alpha + z = 2 
\eea

{\noindent} 
and interestingly enough, the general tendency of the system is to flow towards a short-ranged fixed point (the
long-ranged fixed point comes out to be unphysical with the
specific parameter values, for our particular case). This effect,
as we will see, holds sway in 1+1 dimensions too, where the system
flows towards the KPZ fixed point.

Combining the last two equations, we get

\bea 
\alpha = -K 
\eea

{\noindent} 
Thus the critical exponents come out as

\ber 
z &=& \frac{22}{23} = 0.96 \nonumber \\ \alpha &=&
\frac{24}{23} = 1.04 \nonumber \\ \beta &=& \frac{\alpha}{z}
\approx 1.08 
\eer

{\noindent} 
{\it i. e.} $ \alpha \approx \beta \approx z \approx 1 $
in reasonable agreement with experimental and numerical findings
[14,18,19] (experimental values are: $ \alpha = 0.96 \pm 0.06 $, $
\beta = 0.91 \pm 0.03 $ and $ z = 1.05 \pm 0.08 $), within
experimental error bars. The fact that the theory (and also experiment [18]) predicts $ \alpha \approx 1 $ indicates that the effects of overhangs might be marginal (pg. 110, ref.[1]). Also to be noted is the invariance of the Galilean identity $ \alpha + z = 2 $. Before concluding this portion, it must be mentioned that for the opposite scenario, {\it i. e.} growth under first-order thermal
reemission, an identical analysis as above shows immediately that now the reduced dynamical equation has a form nearly the same as that in eqn.(10) but with a negative non-local potential. This automatically suggests that due to the attractive nature of this potential, the growth finally stops at sufficiently large times ("smoothens") and $ \beta \approx 0 $ [18]. Interestingly, we find that even without thermal reemission, this marked change in the scaling properties, depending on whether it is a growth or an etching process has been discussed elsewhere [25] also.

For the 1+1 dimensional case, we
follow exactly similar lines, the only modification being the
consideration of $ {\theta}' = 0 $ and $ \theta = o $ or $ \pi $
(depending on growth or decay, respectively) in eqn.(6).
Thereafter, proceeding likewise, the dominating long-ranged part
comes out to be $ v(r) \int_{0}^{L} dr' {({\partial}_{r'} h)}^2 $,
with $ v(r) \approx \frac{L}{r} $. Thus in the large time limit, as $
r \rightarrow L $, we see that the system approaches the
conventional KPZ fixed point and naturally the exponents too
resemble the KPZ universality class, which can be looked upon as sort of an analogy with the shadowing case [7]. To avoid unnecessary
repetition of identical calculations, as in the 2+1 dimensional
case, we have neglected any further details in 1+1 dimensions.

All said and done, however, there is still one open question which
needs to be resolved. This is the fact that inspite of both the
available short-ranged and long-ranged fixed points in the 2+1
dimensional case, the system chooses the short-ranged fixed point
(an alternative statement that there is Galilean invariance in the
system, since the other fixed point basically gives an unphysical
picture with $ \alpha < 0 $) although the shadowing effect
fundamentally remains a non-local contribution. This seems to
suggest that whenever we are talking about non-local interactions,
it does not necessarily mean that the long-ranged structure should
control the associated dynamics. Instead the short-ranged part of
the contribution might also take the upper hand, though, obviously
depending on the type of interaction we are considering. The issue
seems to demand further studies. As an adjoinder, we would like to
mention that the 1+1 dimensional situation, being basically
dominated by the KPZ fixed point, no such complexity arises over
there.

The author (A.K.C.) would sincerely like to acknowledge illuminating interactions with Y. -P. Zhao and J. Drotar. All discussions with Dr. Sergei Egorov and Y-J. Lee are acknowledged. A.K.C. is indebted to M.J. Alava for his hospitality at the HUT. 

      *   *   *
$ {}^{*} $Present address of the author: Max-Planck-Institute for the Physics of Complex Systems, N\"othnitzer Str. 38, Dresden, D-01187, Germany. \\
email: akc@mpipks-dresden.mpg.de

\end{document}

%% file: shadow4.bbl
\begin{references}
\bibitem{23} From simple power counts from both sides of the self-energy equation, $ \eta = 1 $, since, otherwise, the right side of the equation becomes a function of two variables, as opposed to the left side, which is only dependent on the external momentum.
\bibitem{24} Putting $ d = 1 $ and $ \rho = 1 $ in the results of [22]. Here it should be remembered that although we are still in the two dimensional plane, the structure of our non-local part features a one dimensional integration only. Hence $ d = 1 $.
\end{references}
